\documentclass{sig-alternate}
\usepackage{multirow}
\usepackage{array}
\usepackage{graphicx}
\usepackage{float}
\usepackage{url}

\begin{document}
\makeatletter
\def\@copyrightspace{\relax}
\makeatother

\title{Science and Ethnicity: How Ethnicities Shape the Evolution of Computer Science Research Community}

\numberofauthors{1}
\author{
\alignauthor Zhaohui Wu$^{\dag}$, Dayu Yuan$^{\ast}$, Pucktada Treeratpituk$^{\ddag}$,  C. Lee Giles$^{\ddag\dag}$\\
\affaddr{$^{\dag}$Computer Science and Engineering, $^{\ddag}$Information Sciences and Technology}\\
\affaddr{Pennsylvania State University, University Park, PA, USA}\\
\affaddr{$^{\ast}$Google, Mountain View, CA, USA}\\
\email{laowuz@gmail.com, dayuyuan@google. com, \{pxt162, giles\}@ist.psu.edu}
}

\maketitle
\begin{abstract}
Globalization and the world wide web has resulted in academia and science being an international and multicultural community forged by researchers and scientists with different ethnicities. How ethnicity shapes the evolution of membership, status and interactions of the scientific community, however, is not well understood. This is due to the difficulty of ethnicity identification at the large scale. We use name ethnicity classification as an indicator of ethnicity. Based on automatic name ethnicity classification of 1.7+ million authors gathered from Web, the name ethnicity of computer science scholars is investigated by population size, publication contribution and collaboration strength. By showing the evolution of name ethnicity from 1936 to 2010, we discover that ethnicity diversity has increased significantly over time and that different research communities in certain publication venues have different ethnicity compositions. We notice a clear rise in the number of Asian name ethnicities in papers.  Their fraction of publication contribution increases from approximately $10\%$ to near $50\%$ Yearly accumulated population size of from 1970 to 2010. We also find that name ethnicity acts as a homophily factor on coauthor networks, shaping the formation of coauthorship as well as evolution of research communities.
\end{abstract}

\section{Introduction}
``\textit{Science knows no country, because knowledge belongs to humanity, and is the torch which illuminates the world.}''
\begin{flushright}
 ---Louis Pasteur.
\end{flushright}

The word ethnicity is derived from the Greek word, `ethnos', which means a nation or tribe~\cite{Bhopal:Glossary}. While various definitions for ethnicity have been proposed, there is no consensus as to its precise meaning. In general, most definitions imply one or more of the following features: common geographic origins; shared traditions and culture that are maintained between generations leading to a sense of identity; and common language or religious faiths~\cite{Smith:Ethnicity}. Generally,  a good educated guess of a person's ethnicity can be based purely on their name, especially for ethnicities one is familiar with. Chinese can usually identify other Chinese just on their names. Ethnicity is an important demographic indicator used in various applications including marketing~\cite{Holland:Ethnic}, public policy~\cite{Mays:Classification}, funding rewards~\cite{eth:11}, epidemiology~\cite{Senior:Ethnicity} and biomedical research~\cite{Gill:Limitations}.

Over the past decades, research and scholarship have become quite diverse with a growth in scholars from several different countries and ethnicities.
To our knowledge, how ethnicity shapes the evolution of science has not been studied. Questions of interest could be whether research is becoming more diverse? Is ethnicity a homophily factor~\cite{Miller:Birds} that affects research collaboration and community formation? Are there any collaboration preferences among different ethnicities? If yes, then ``ethnicity" could be an important factor for better understanding scientific collaboration~\cite{MEJ:Coauthorships,Collaborationincomputerscience,Collaboration:Jian}, social ties and influence~\cite{SocialTie,Socialinfluenceanalysis}, and communities~\cite{Groupformation,Community:Deng}.

\begin{figure}
\centering
\vspace{-2.3cm}
\includegraphics[width=0.5\textwidth]{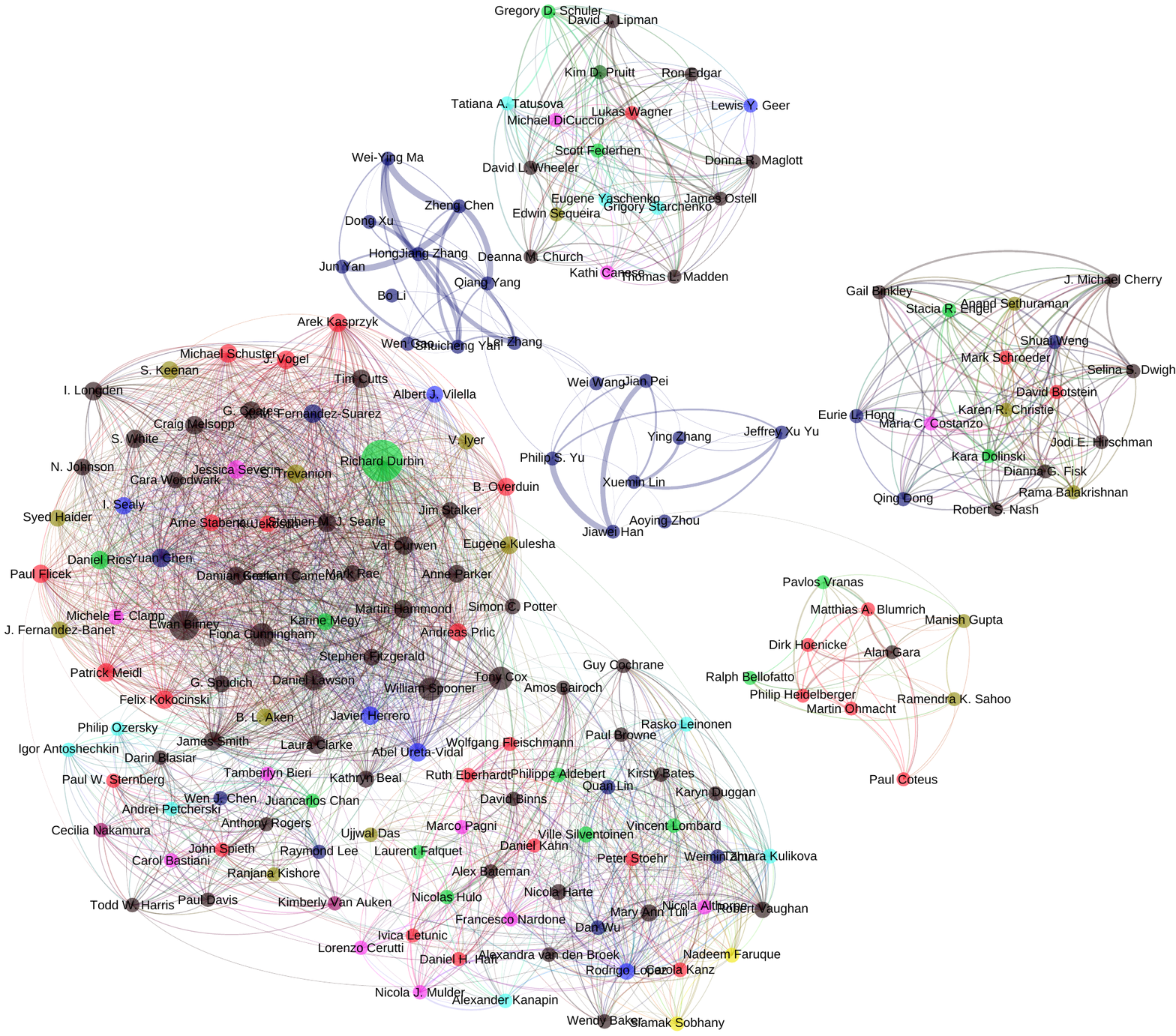}
\vspace{-3cm}
\caption{Name ethnicity of a coauthor network of the top coauthored 173 researchers in DBLP}
\label{fig-eth-173}
\end{figure}

As an example, Figure~\ref{fig-eth-173} shows the clusters of a coauthor network formed by 173 researchers with the most number of coauthors. Different colors represent different name ethnicities while the color of an edge is the average mixed color of its two nodes. The size of a node is proportional to its degree. Most nodes in the largest cluster are black (English) and red (German) while the two smaller clusters are all navy (Chinese) nodes. This hints that ``ethnicity" may matter in coauthorship relationships.

While studies of science at the country level ~\cite{Luukkonen:Understanding,W:National} always focus on investigating statistics and patterns of international scientific collaboration, we argue the exploration at the ethnicity level could potentially provide additional insight for policy in research funding and distribution~\cite{eth:11}  and education recruitment, especially for immigration oriented countries with multiple ethnicities such as the USA. The rich and well-archived scholarly and scientific repositories on Web, such as arXiv, DBLP, CiteseerX, Arnetminer, and Medline/Pubmed\footnote{\url{http://en.wikipedia.org/wiki/List_of_academic_databases_and_search_engines}}, provide  opportunities to understand the evolution of name ethnicity from an international and multicultural perspective. To systematically explore such questions, we need a large sample of researcher names and publications, and a method to identify the ethnicity of each name. As such we leverage some of the online databases previously mentioned, which contain large scale author profiles and publication records in computer science. Despite the lack of accurate ethnicity information, we believe authors' ethnicity can be inferred solely based on their names using name ethnicity classification from the U.S. Census Bureau's dataset~\cite{ePluribus} or an open name source such as Wikipedia~\cite{Ambekar:opensources,Pucktada:name}. We also make the assumption that name ethnicity is an indicator of ethnicity in general. Though we feel that is the case in general, there will be exceptions.

Based on the syntactic knowledge residing in the sequences of alphabets and phonetic sounds within a name, we use 215,672 personal names with identified ethnicity crawled from Wikipedia to train a 12 class name ethnicity classifier based on multinomial logistic regression. This classifier  can infer ethnicities from names with an accuracy as 85\% on average~\cite{Pucktada:name}. The 12 classes are representative of the  population size of world ethnicities. However, these classes could be extended given more labeled ethnicity data. Note that an ethnicity of an author does not imply his/her nationality. For example, authors of English, German, and Chinese ethnicity might all have the same nationality. By applying our classifier to 1,636,790 author names from Arnetminer together with all author names in 1,632,442 publication records from DBLP, we obtain the ethnicity of all authors and conduct a systematic study of the evolution of ethnicity in computer science research based on various characteristics such as population dynamics and scientific contribution. We summarize our main findings below:
\begin{itemize}
\item We observe a rising trend in Asian ethnicities in population size, scientific output, and collaboration strength. Specifically, Chinese and Indian names on scholarly papers are rising with respect to European names, especially  after the 1990s.
\item Different ethnicities have different trends in author population size and publication output not only for the whole of research, but also in different research communities. In particular, we found Asian ethnicities are playing a stronger role in the Data Mining and Information Retrieval communities while European ethnicities maintain their strong presence in the Artificial Intelligence and Algorithm and Theory communities.
\item Ethnicity is shown to be a strong homophily factor in a coauthorship network, which influences collaboration and further shapes the formation of communities in a scholarly or academic social network. Surprisingly, we found that nearly 50\% of coauthor collaboration clusters (with \#nodes>=10) have no less than 50\% nodes from a single ethnicity.
\end{itemize}
The rest of the paper is organized as follows. Section 2 introduces the name ethnicity classification technique using Wikipedia. Section 3 describes the open datasets for our study. Section 4 studies name ethnicity characteristics in computer science based on population dynamics and scientific contribution, while section 5 studies the collaboration preference and strength among name ethnicities. Section 6 discusses the related work about name ethnicity identification and its applications.  Section 7 ends with conclusion and future work.

\section{Name-Ethnicity Classification}
Defining ethnicity classes is difficult because of the subjective and changing nature of ethnic identification. There is no consensus on what constitutes an ethnicity~\cite{Ethnicstatistics}.
We first briefly introduce our name ethnicity classifier trained using Web data. More technical details can be found in previous work ~\cite{Pucktada:name}.

We take advantage of Wikipedia and their categories as a source for collecting personal names of different ethnicities.
First, for each target ethnicity \emph{n}, we pick the Wikipedia category \emph{n\_people} as the root node and then employ BFS to transverse all subcategories and pages reachable within depth of 4. Simple filtering heuristics is used to restrict the link transversals so that only personal names are kept. For example, we only include subcategories whose titles contain the word `\emph{people}' or `\emph{s of}' (e.g. `\emph{Members of the Institut de France}') or end with plural nouns (e.g. `\emph{entertainers}'). The leaf nodes resulting from the transversal are then collected as personal names from the nationality. Note that neither our heuristics nor Wikipedia categories are perfect. For instance, names under `British people of Indian descent', which might be of Indian ethnicity, will be filed under English names. Non-personal names could also be included. For instance, musical group names such as `\emph{Spice Girls}' is included because it is under `British musicians'. As such, we also manually curate the resulting name lists, removing as best we could any such obvious misassigned names.

We automatically harvest a total of 215,672 personal names (after curation) from 19 nationalities, which are then grouped into 12 ethnic groups as shown in Table~\ref{tab-eth-category}. Personal names of Egyptian, Iraqi, Iranian, Lebanese, Syrian and Tunisian are for convenience grouped together as Arabic names (ARA) though we know that Iranian names are different. Names of Spanish, Columbian, Venezuelan are grouped together as Spanish names (SPA). To identify the ethnicity of a name, instead of analyzing each name component (e.g. first name, middle, and last name) as a unit as is typically done, we encode a full name as sequences of characters and phonemes, based on the hypothesis that names of different ethnicities have identifiable sequences of alphabets and phonetics. We then train a multinomial logistic regression classifier using four types of features including non-ASCII, character ngrams, double metaphone ngrams~\cite{Philips:2000} and Soundex~\cite{Knuth,mortimer1995soundex}.

To assess the performance of our name-ethnicity classifier, we randomly split the name list collected from Wikipedia into 70\% training data and 30\% test data, where each full name is a training or testing instance. The results of each class are shown in Table \ref{tab-eth-category}. The overall classification accuracy is 0.85, with Japanese as the most identifiable name ethnicity (F1=0.96), followed by Korean (F1=0.92) and Arabic names as the worst (F1=0.79). In general, the classifier does well identifying East Asian names (CHI, JAP, KOR, and VIE) with over 90\% precision and recall with just one exception, VIE's recall. Overall, our name ethnicity classifier can infer ethnicity from names with fairly high accuracy. Our result is comparable overall to other work ~\cite{Ambekar:opensources}, and is significantly better at identifying some ethnic groups such as German and East Asian names. The demo of our ethnicity classification can be found on line.\footnote{\url{ http://singularity.ist.psu.edu/ethnicity}} Given a name, our ethnicity classifier will output the top three candidate ethnicities along with the confidence score. We choose the first rank ethnicity with confidence score larger than 1/3 as the name's ethnicity, otherwise we label the name's ethnicity as other (OTH).
We plan to make our name ethnicity classifier publicly available in the near future.

\begin{table}
\centering
\caption{\label{tab-eth-category}Name ethnicity with its corresponding categories from Wikipedia and the precision, recall and F1 scores}
\vspace{2mm}
\scalebox{0.82}{
\begin{tabular}{l|l|m{4cm}|c|c|c} \hline
Eth. &\#Names&Categories&Pre.&Rec.&F1\\\hline \hline
ENG &28,624&British&0.79&0.85&0.82\\
GER &35,101&German&0.84&0.85&0.85\\
FRN &29,271&French&0.80&0.80&0.80\\
SPA &15,154& Columbian, Venezuelan, Spanish&0.82&0.79&0.81\\
RUS &19,580& Russian&0.90&0.85&0.87\\
ITA &23,328& Italian&0.85&0.86&0.85\\
IND &21,271& Indian&0.89&0.86&0.87\\
CHI &10,385& Chinese&0.92&0.90&0.91\\
JAP &17,790& Japanese&0.97&0.95&0.96\\
KOR&3,750& Korean&0.93&0.92&0.92\\
VIE &859& Vietnamese&0.93&0.83&0.88\\
ARA &10,559& Egyptian, Iranian, Iraqi, Lebanese, Syrian, Tunisian&0.79&0.78&0.79\\ \hline\hline
\end{tabular}}
\end{table}

\section{Datasets}
Ideally, to study ethnicity in computer science authorship comprehensively, we need to collect all the scholars' names and publications for each year. While these could not be totally harvested, the well-archived scholarly and scientific databases and repositories on Web provide us with very representative resources. Specifically, we use the author list from Arnetminer and publication records from DBLP. The Arnetminer author list contains 1,636,790 unique author names gathered from its research profiles, with research interest, homepage URLs.\footnote{\url{http://arnetminer.org/}} The DBLP dataset contains 1,632,442 publication records in computer science from 1936 to 2011\footnote{\url{http://arnetminer.org/DBLP_Citation}}~\cite{Jie:ArnetMiner}, with 858,765 unique author names appearing among them. All records have metadata including title, author list, publication year and venue information. These author names with their yearly publication records are used to study the evolution of population and publication contribution of each ethnicity. Venue information such as conference name is used for defining a research community. The union of the two set of author names includes 1,790,266 names in total, which is used to study the overall population distribution among different name ethnicities.
In addition, we also use a list of h indices for computer scientists that contains 745 researchers whose h index >= 40.\footnote{\url{http://www.cs.ucla.edu/~palsberg/h-number.html}} This provides a sample for studying ethnicity on well published researchers.
\begin{figure}
\centering\vspace{-3mm}
\includegraphics[width=0.5\textwidth]{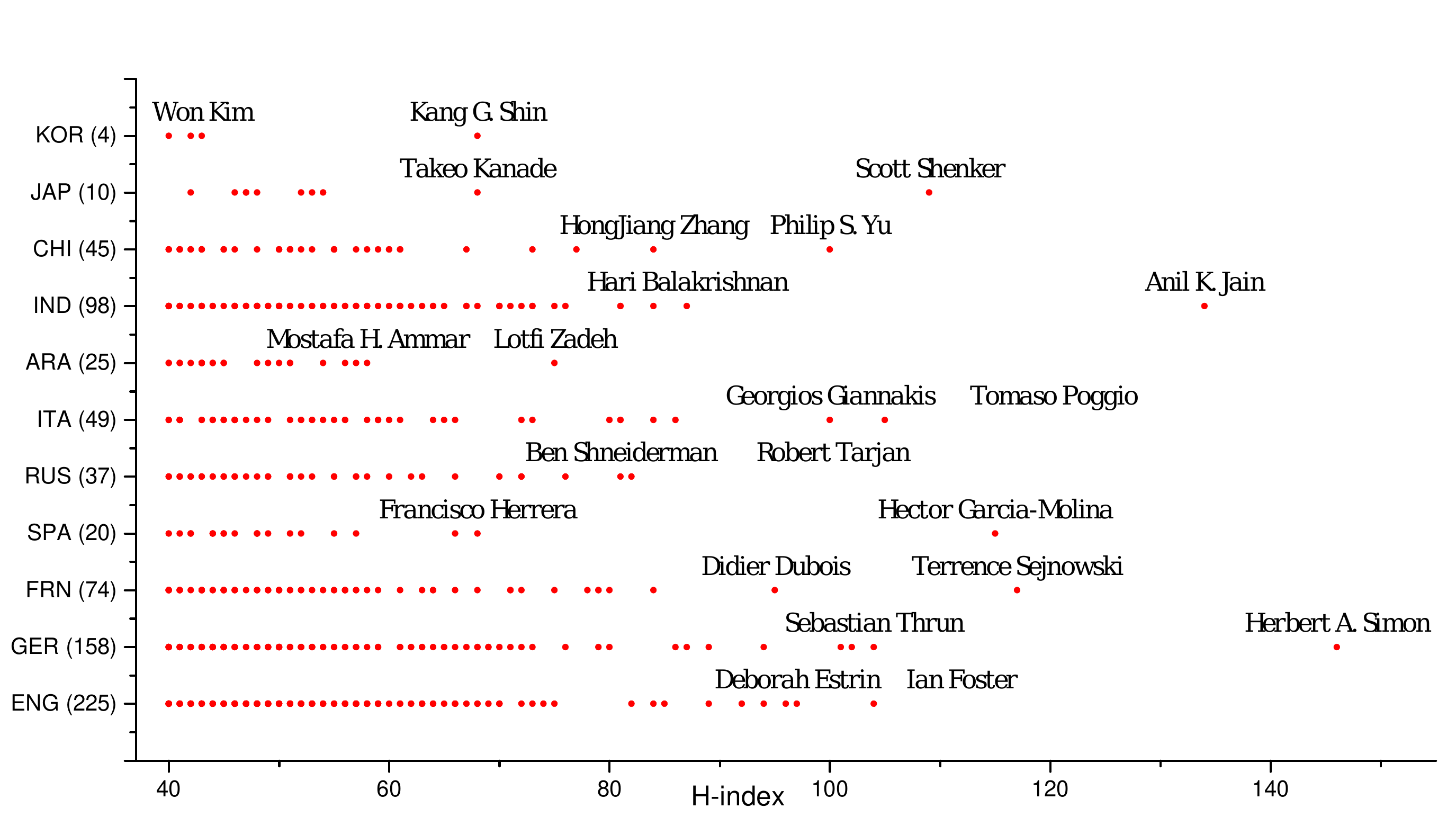}
\vspace{-6mm}
\caption{Name ethnicity of 745 computer researchers whose h-index >=40. The y-axis indicates the ethnicity and the number of scholars belonging to that name ethnicity}
\label{fig-eth-h-index}
\end{figure}
\begin{figure}
\centering
\includegraphics[width=0.5\textwidth]{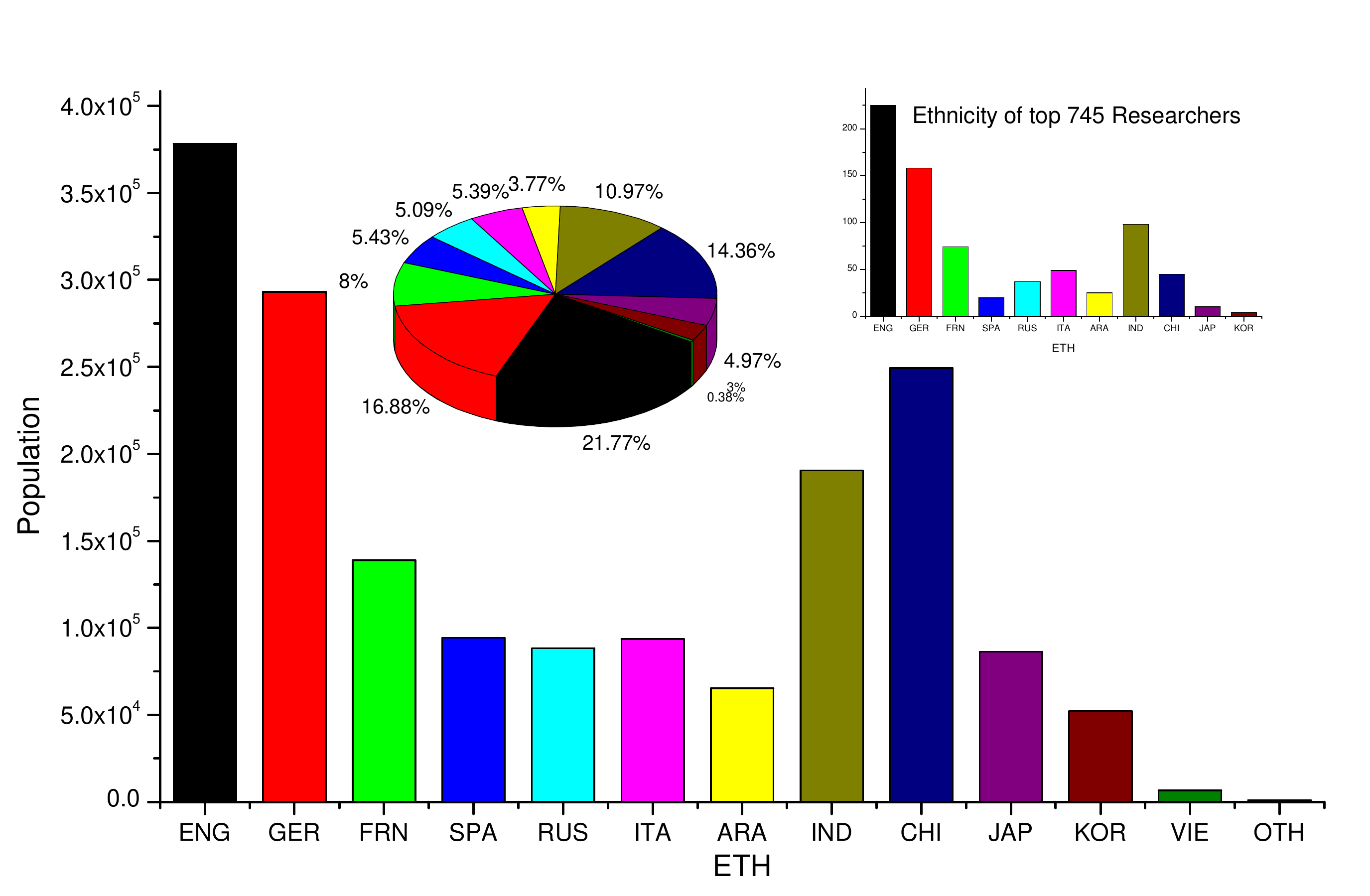}\vspace{-3mm}
\caption{Population of different name ethnicities in computer science based on author names from the union set of Arnetminer and DBLP (the pie chart shows the fractions in percentage). The upright sub-figure shows the population of ethnicities on the 745 top h-index researchers.}
\label{fig-eth-dis}
\end{figure}
\section{\label{Eth-CS}Name Ethnicity of Computer Science Family}
We first explore the name ethnicity composition of the whole of computer science in terms of population statistics. We show the population distribution of top computer scientists with h-index >= 40 in Section~\ref{h-index}, after which, the population distribution and growing status of different ethnicities overall are shown in Section~\ref{sec-pop-dis} and Section~\ref{sec-pop-dyn} respectively.
In Section~\ref{output} we measure the publication contribution and how it changes with each name ethnicity.
Finally, we explore in Section~\ref{eth-conf} the name ethnicity statistics and how they evolve in different research domains such as theory and data mining.

\begin{figure}
\centering
\hspace{-10pt}\vspace{-10pt}
\includegraphics[width=0.49\textwidth]{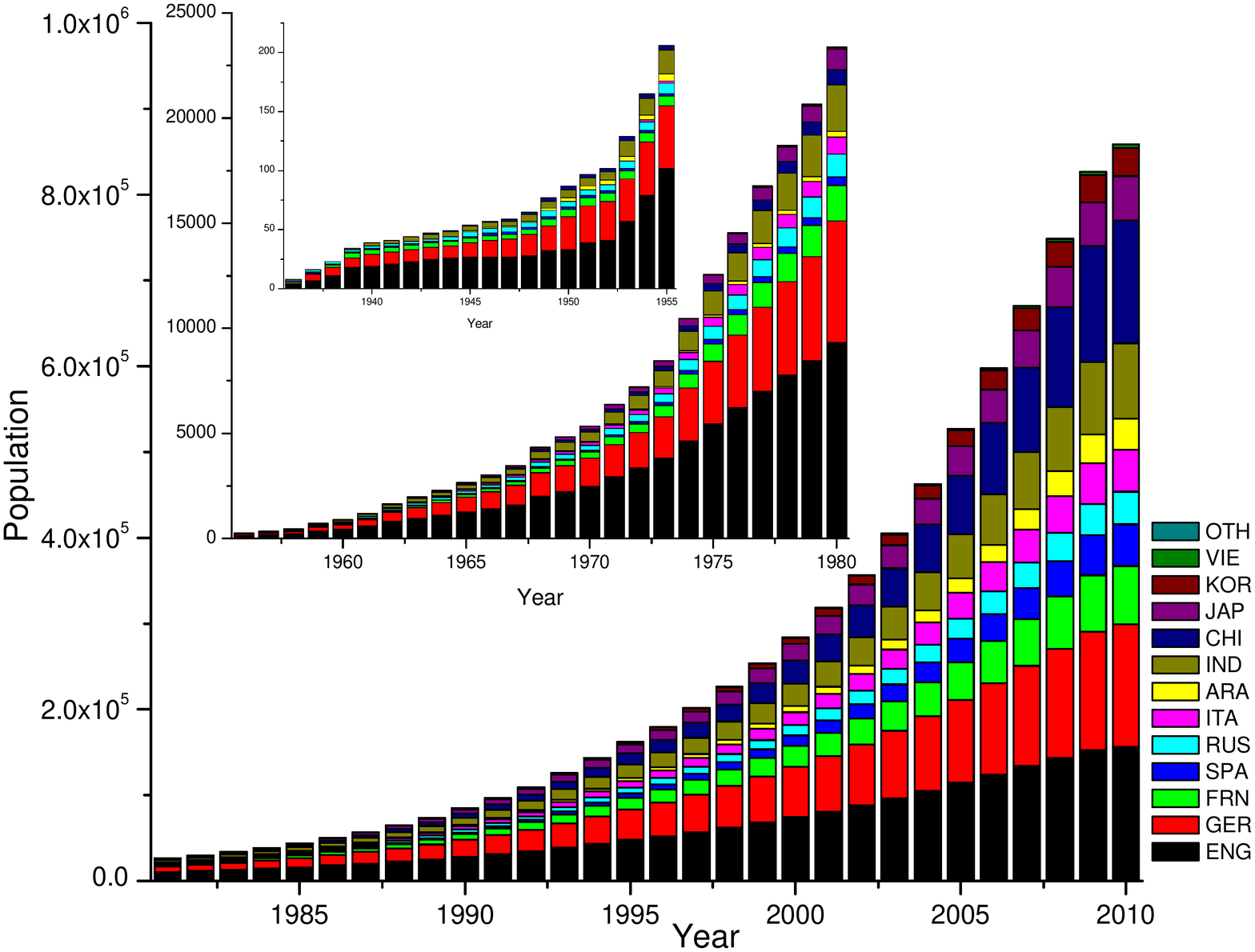}
\vspace{-15pt}
\caption{Accumulated population of different name ethnicities from 1936 to 2010 based on author names from DBLP publication records.}
\label{fig-eth-pop}
\end{figure}

\subsection{\label{h-index}Name Ethnicity and H Index}
The h-index, defined as the number of papers with citation number higher than or equal to $h$, has been considered as a popular index to characterize the total scientific impact of a researcher~\cite{Hirsch:Hindex}. Regarding the name ethnicities, one question to consider is how many researchers are from a specific ethnicity, say Chinese or Indian, with an h-index beyond 40. Figure~\ref{fig-eth-h-index} shows the scatter plot of all the 745 researchers in the space of name ethnicity and h-index, with the exact names of top 2 researchers of each name ethnicity. Note that ENG and GER together dominate 51.4\% of the total number of top researchers, while Chinese and Indian together only 19.2\%. It is reasonable that the top scientist list is dominated by European names, since most of them are pioneers in the early days of computer science. We believe as $h$ goes smaller, there would be more researchers with Asian name ethnicities such as Chinese.  

\subsection{\label{sec-pop-dis}Population Distribution}
An intuitive speculation would be that the number of top researchers in a name ethnicity is proportional to its overall total number. The population distribution among name ethnicities in the entire 1.6+ million names is thus examined for the name ethnicity composition of the top researchers. Figure~\ref{fig-eth-dis} depicts the number as well as the percentage of each name ethnicity's population, showing a similar composition compared to the upright subplot of top 745 researchers. ENG and GER dominate the population as the top two ethnicities, while CHI and IND follow as the third and fourth. It is interesting to note that CHI has a larger population than IND but a much smaller number of  top researchers. Also, a similar phenomenon occurs in SPA, which has slightly larger population than RUS and ITA but a smaller number of top researchers. We hypothesize that this is due to the relatively smaller fraction of CHI and SPA in earlier years, but faster growing in more recent years. By checking the accumulated population size at 1980 (see Figure~\ref{fig-eth-pop}), we see that CHI has a smaller number than IND and SPA has a smaller number than either RUS or ITA. We will further address this problem in the next Section based on more analytic results.

\subsection{\label{sec-pop-dyn}Evolution of Population}
\begin{figure}
\centering\vspace{-2mm}
\includegraphics[width=0.5\textwidth]{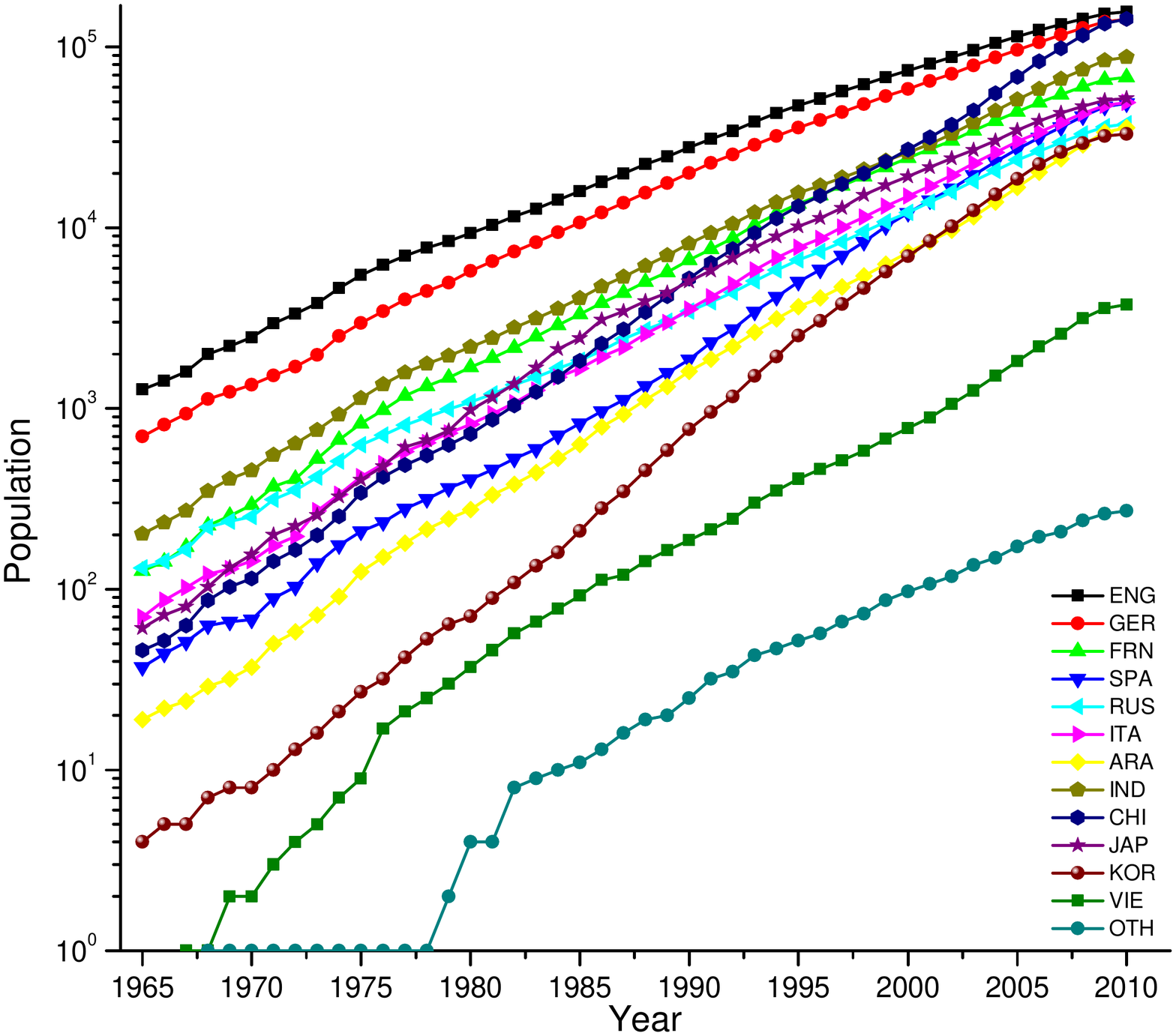}\vspace{-4mm}
\caption{Yearly accumulated population size of different name ethnicities based on DBLP records.}
\label{fig-eth-acc}
\end{figure}

\begin{figure*}
\centering
\includegraphics[width=0.95\textwidth]{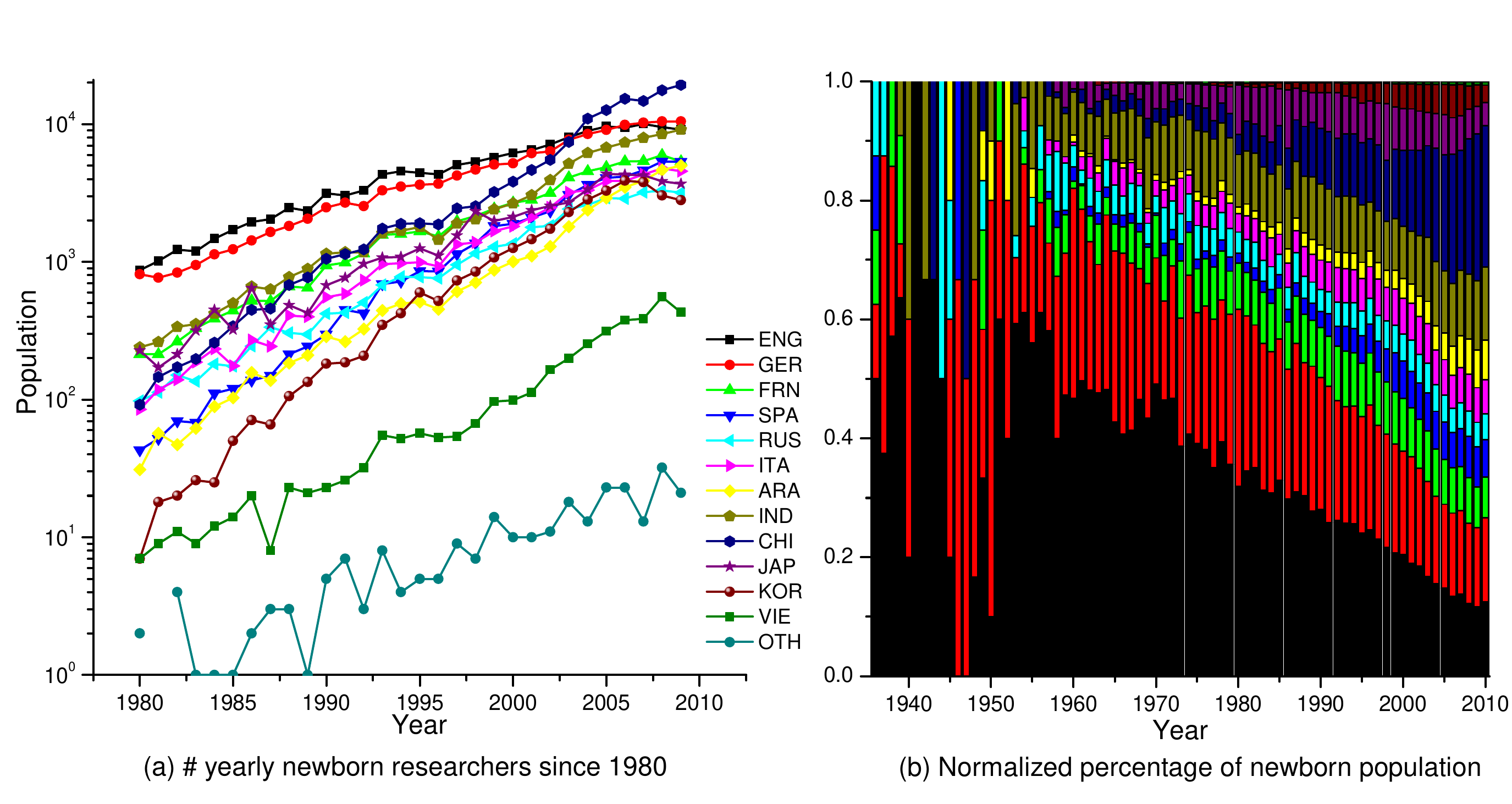}
\vspace{-3mm}\caption{Yearly new author population size of different ethnic groups based on DBLP publication records. The left shows the change of absolute number while the right shows the normalized percentage by the length of bars.}
\label{fig-eth-new}
\end{figure*}

To better understand the evolution of population size of different ethnicities, we present the accumulated population size of each name ethnicity from 1936 to 2010 in Figure~\ref{fig-eth-pop} (with a log scale plot from 1965 to 2010 in Figure~\ref{fig-eth-acc}), and the yearly new author population size since 1980 in Figure~\ref{fig-eth-new} (with the absolute number in the left sub-figure and the normalized percentage to the right ). For each year $t\in [1936, 2010]$, the accumulated population size is calculated by counting the number of unique author names (assuming disambiguation) that appear in publications between $[1936, t]$, while the new author population size in $t$ is the number of new author names in publications of year $t$.
We contend that the two figures, to some extent, give insight into the number disparities discussed in Section~\ref{sec-pop-dis}. Only after 2000 does the CHI name ethnicity began to catch up with IND in population size and SPA surpassed RUS and ITA even later. The population distribution in 1990s is much closer to that of top h-index researchers today, suggesting that today's output (in terms of number of top scientists) depends on the input of 20 years ago.

Figure~\ref{fig-eth-new} (a) suggests that all ethnicities have an exponentially growth trend, which leads to the exponential increase trends shown in Figure~\ref{fig-eth-acc}.
It can be observed that ENG and GER have a strong presence in the entire population with tremendous growth before 1980 but then slowed afterwards. The new author population of each name ethnicity increases approximately exponentially and CHI, KOR and SPA have a relatively larger increased rates. In 2004, CHI's new author population has surpassed ENG and GER, now being the top one for all ethnicities.
Figure~\ref{fig-eth-new} (b) gives more insights in the difference of growing trends among ethnicities. The fraction of new author population in ENG and GER are dramatically decreasing compared to others, due to the slower growth rate.
If taking the border of the yellow bar (ARA) and magenta bar (ITA) as the boundary of Asian and European ethnicities, we can clearly see Asian ethnicities are growing much faster than European. It is also worth noting that SPA is the only European name ethnicity that have an increasing growth rate after 1990s.

Though we have no statistics on age, based on the above figures we may infer that ENG and GER have a slower population growth while CHI and IND  keep growing due to already large population bases.
The population growth of an name ethnicity can be modeled by the Logistic population model~\cite{Sharon95}
\begin{equation}
P(t)=\frac{P_m}{1+(P_m/P_0-1)e^{-r(t-t0)}}
\end{equation}
where $P(t)$ is the population size at year $t$, $P_0$ is the initial population size at year $t_0$, $P_m$ is the carrying capacity (the maximum population size of research support for the name ethnicity), and $r$ is the population growth rate. By applying the model to different ethnicities, we found they are in different periods of their growth curves: most European ethnicities, JAP and KOR are close to the inflection point; CHI, IND and VIE are far ahead it, indicating more growth potential in the future.

\subsection{\label{output}Name Ethnicity of Scientific Output}
\begin{figure*}
\centering
\includegraphics[width=0.95\textwidth]{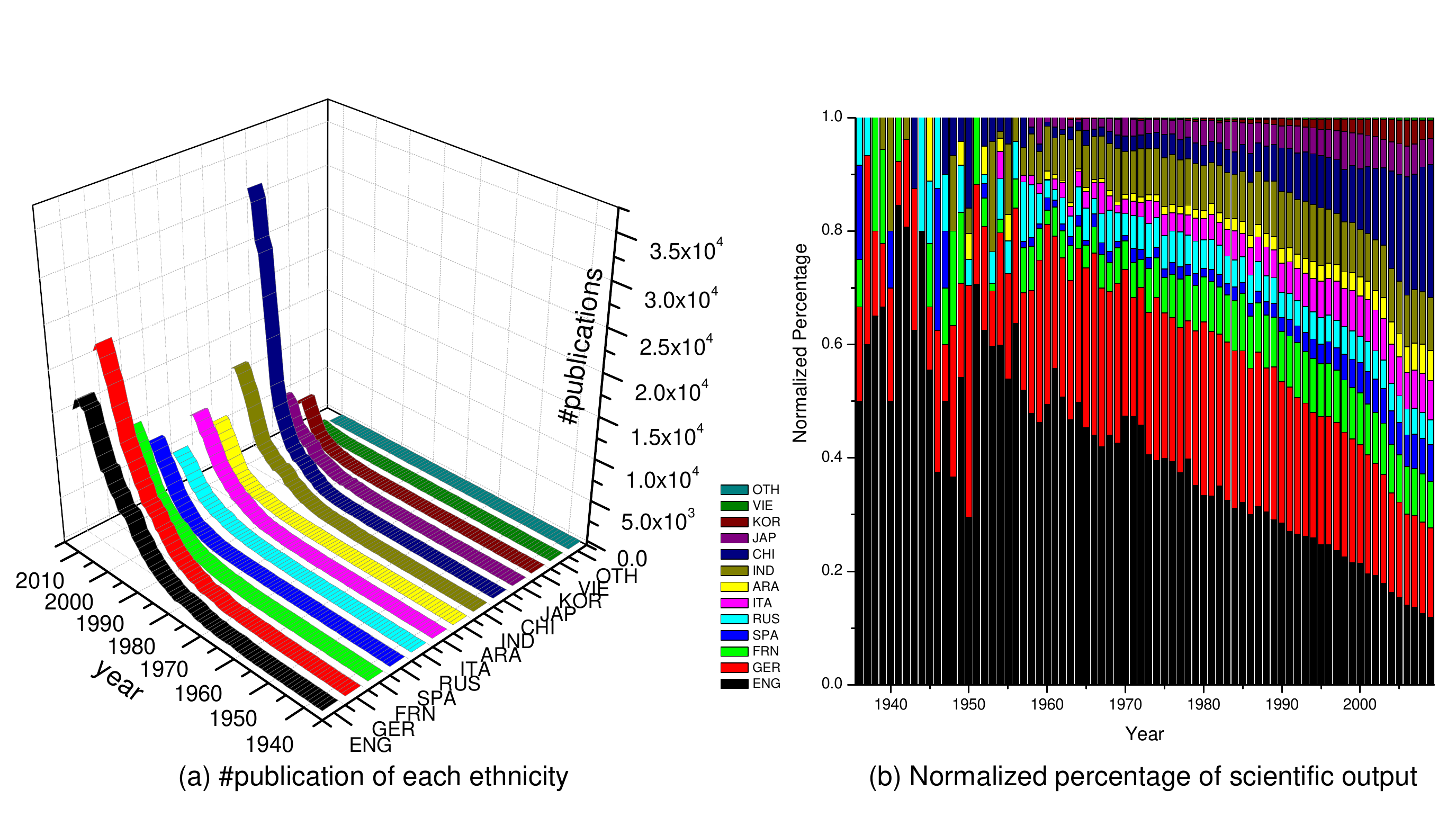}
\vspace{-7mm}
\caption{Publication contribution of each name ethnicity based on number of publications. The left shows the change of absolute number while the right shows the normalized percentage.}
\label{fig-single-eth-contrib}
\end{figure*}

Our findings indicate a larger population is positively correlated to more top h-index researchers in the future. Another question to explore would be ``will the larger researcher population lead to a larger scientific output in the same period?''. To answer this question, we measure the scientific output of different ethnicities based on number of publications. For a paper of $K$ coauthors, $1/K$ paper will be counted for each author. Figure~\ref{fig-single-eth-contrib} (a) shows the number of publications of each name ethnicity every year while Figure~\ref{fig-single-eth-contrib} (b) shows the corresponding normalized percentage by the length of bars with different colors.

By comparing Figure~\ref{fig-single-eth-contrib} (b) with Figure~\ref{fig-eth-new} (b), we could conclude that the publication output in a year is positively correlated to the new author population size in that year. In other words, new scholarly output depends on the number of new researchers.

We found that ENG tops all the others in scientific output before 2001. GER has surpassed ENG slightly since 2001. However,  CHI has taken the top position from GER since 2005. Considering the recent advantage of CHI for yearly new authors, it is reasonable to believe CHI will continue  to increase in the near future.

The boundary between Asian and European ethnicities in Figure~\ref{fig-single-eth-contrib} (b) again gives a better Asian and European comparison. From 1970 to 2010, the Asian fraction has increased nearly 40\%, from around 10\% to near 50\%, mostly contributed by the increase of CHI (around 20\%). It is worth noting that this near 40\% happens to be close to the decreased share of ENG together with GER, from around 70\% to around 30\%, which indicates the other European ethnicities' scholarly output proportions have been kept relatively constant.

\subsection{\label{eth-conf}Name Ethnicity in Research Communities}
The above analysis regards computer science as a whole with the topics of different research communities not explored. How does the name ethnicity composition evolve in different communities? Does the ``European-Asian shift'' effect occur differently in different communities? To answer these questions, we examine the ``impact'' of each name ethnicity to a research community based on its number of authors in the community. For simplicity, here a community is represented by a set of popular conferences in the same area, as listed in Table~\ref{tab-confs}, where four communities, information retrieval (IR), data mining (DM), artificial intelligence (AI) and algorithms and theory (AT) are selected for study as cases.

\begin{table}
\centering
\caption{\label{tab-confs}Communities and selected conferences (colors correspond the bubbles colors in Figure~\ref{fig-eth-confs}}
\vspace{2mm}
\scalebox{0.74}{
\begin{tabular}{|l|l|} \hline
Community &Selected Conferences\\\hline \hline
IR (Information Retrieval)  &SIGIR(green), CIKM(black), TREC(blue), \\
&CLEF(red)\\\hline
DM (Data Mining)  &KDD(red), ICDM(black), SDM(Cyan),\\
& PKDD(blue), PAKDD(green)\\\hline
AI (Artificial Intelligence) &IJCAI(blue), AAAI(black), ICML(green), \\
&UAI(magenta), NIPS(Cyan), AAMAS(red)\\\hline
AT (Algorithm and Theory)& STOC(magenta), SODA(Cyan), FOCS(red), \\
&ICALP(green), LICS(blue), CONCUR(black)\\ \hline
\end{tabular}}
\end{table}

\begin{figure*}
\centering
\hspace{-15pt}
\includegraphics[width=0.95\textwidth]{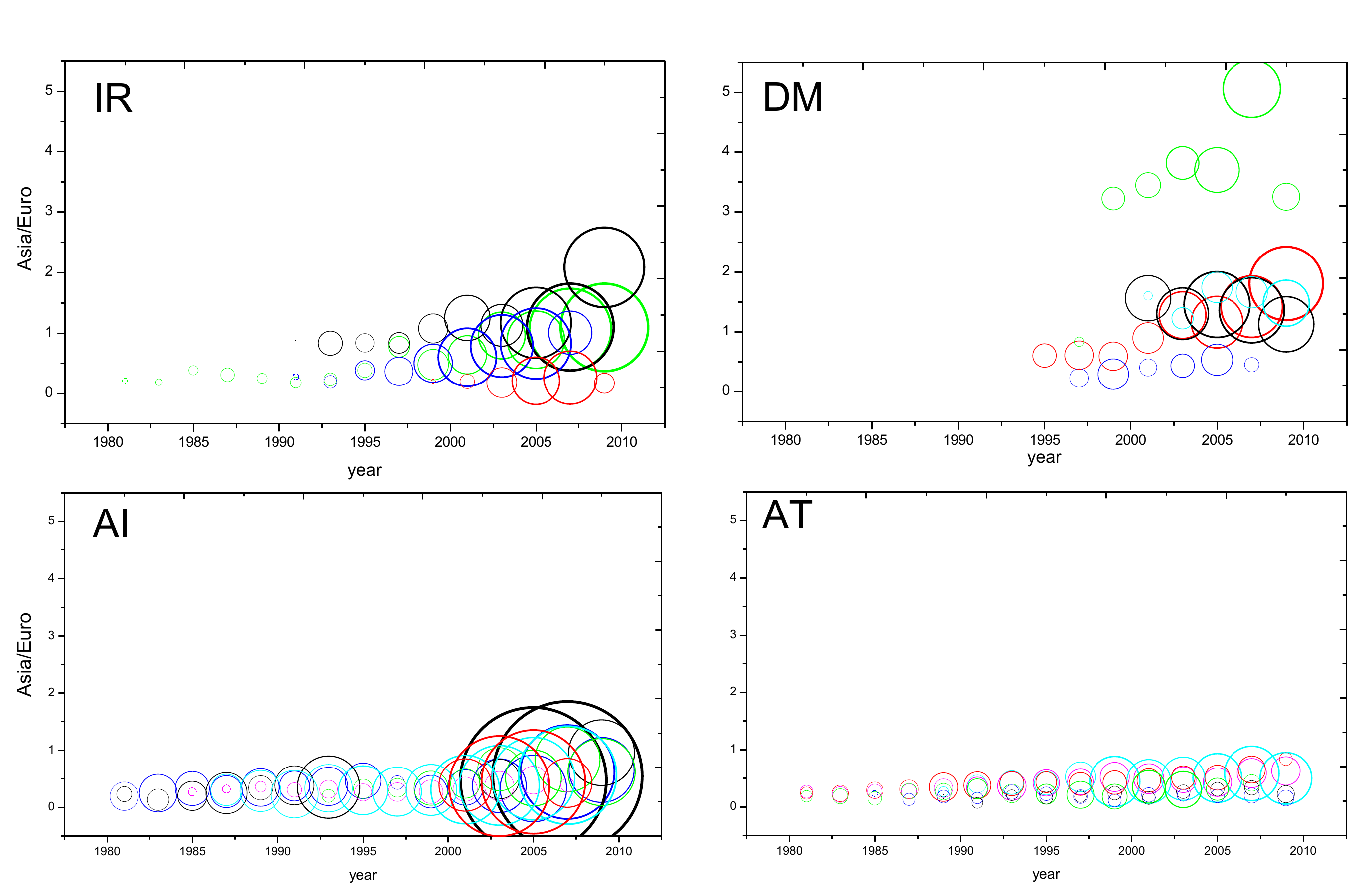}
\vspace{-2mm}
\caption{Ratio of \#Asian names to \#European names in research communities based on conferences}
\label{fig-eth-confs}
\end{figure*}

For each conference in a community, we define its Asian-European ratio as the number of total unique Asian names over the number of total unique European names in the publications of the conference. A similar definition applies to the CHI-ENG ratio. 
The evolution of the ratio in the four communities is shown in Figure~\ref{fig-eth-confs}. Within a community, different conferences are represented by bubbles of different colors. The X-axis and Y-axis values of the bubble center are year and ratio respectively. The size of bubbles indicates the size of a conference, which is proportional to the total number of unique names in the publications of the conference. For example, in IR community, SIGIR is represented by green bubbles. The ratio has become larger than 1 since 2003, with an increasing conference size.

The results clearly show that different communities have unique name ethnicity composition and evolution. The Asian-European ratio keeps growing in the IR and DM communities, indicating that Asian ethnicities are playing a more important role in the two communities. 
Since CHI and ENG are the largest name ethnicity in Asian and European ethnicities respectively, the evolution trends of Asian-European ratio and CHI-ENG ratio looks close on a conference.
The plots also enable a rank order of the conference based on Asian-European ratio. For example, we may contend that in IR community CIKM > SIGIR > TREC > CLEF in terms of Asian-European ratio, indicating that Asian ethnicities publish the most in CIKM while the fewest in CLEF.
However, in AI and AT, the two ratios are less than 1 most the time, implying that European ethnicities are still greater in these two areas, although two ratios are approaching or slightly larger than 1 for IJCAI(blue), AAAI(black), and  ICML(green) in AI in the latest two years, showing an increasing impact of Asian and CHI in these three conferences.
Hence, we could conclude that the ``European-Asian shift'' effect also exists in research communities. DM has the strongest effect, followed by IR and AI, while AT shows as the weakest.

\begin{figure*}
\centering
\hspace{-15pt}
\includegraphics[width=0.95\textwidth]{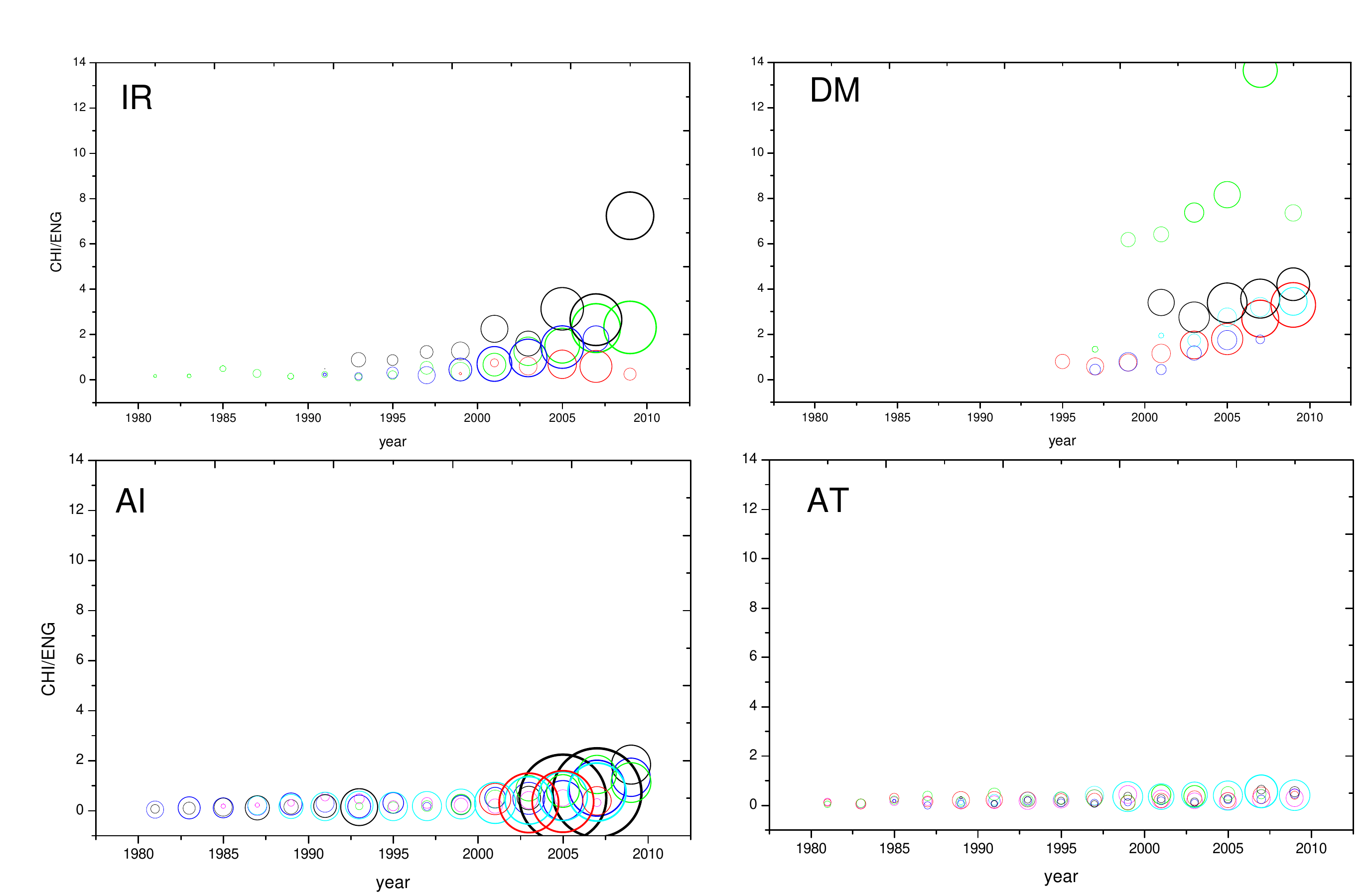}
\caption{Ratio of \#Chinese names to \#English names in research communities based on conferences}
\label{fig-eth-confs-C-E}
\end{figure*}

It is interesting to note that sizes of IR, DM and AI have all increased after 2005 while size of AT has not changed as much, implying the former three areas might be more popular than AT. Besides, we can observe that DM is a relatively younger community than the other three, as those DM conferences appear after 1990s, much later than conferences of other communities.

In addition, we find that the name ethnicity of a conference depends a lot on the conference's location. For example, in DM, PAKDD (green), as an Asian conference, has the highest Asian-European ratio and CHI-ENG ratio while PKDD (blue), as an European conference, has the lowest ratios. However, more study are needed to verify the correlation between a conference's geolocation and name ethnicity.

\section{Name Ethnicity of Scientific Collaboration}
Previous sections revealed the history of different ethnicities mainly based on their population and scientific output. This section studies the collaboration among different ethnicities, where name ethnicity is proven to be a strong homophily factor~\cite{Miller:Birds} on coauthor network. Moreover, the evolution of collaboration patterns can well explain the rise of Asian ethnicities.
\subsection{Name Ethnicity on Coauthor Network}
The small coauthor network showed in Figure~\ref{fig-eth-173} gives us an intuition that name ethnicity could be an important factor to shape research collaboration and form research communities. However, it is still unknown whether it is a local phenomenon for a few certain ethnicities or a common exist on the whole coauthor network. To this end, we build a coauthor network based on the DBLP dataset and run cluster/community detection~\cite{Vincent:Fast} on the largest connected component, which gives us an output of 727 clusters. Since we focus on the name ethnicity and different authors with the same name will definitely the same name ethnicity, name disambiguation is not performed here.

\begin{figure}
\centering
\vspace{-1.8cm}
\includegraphics[width=0.5\textwidth]{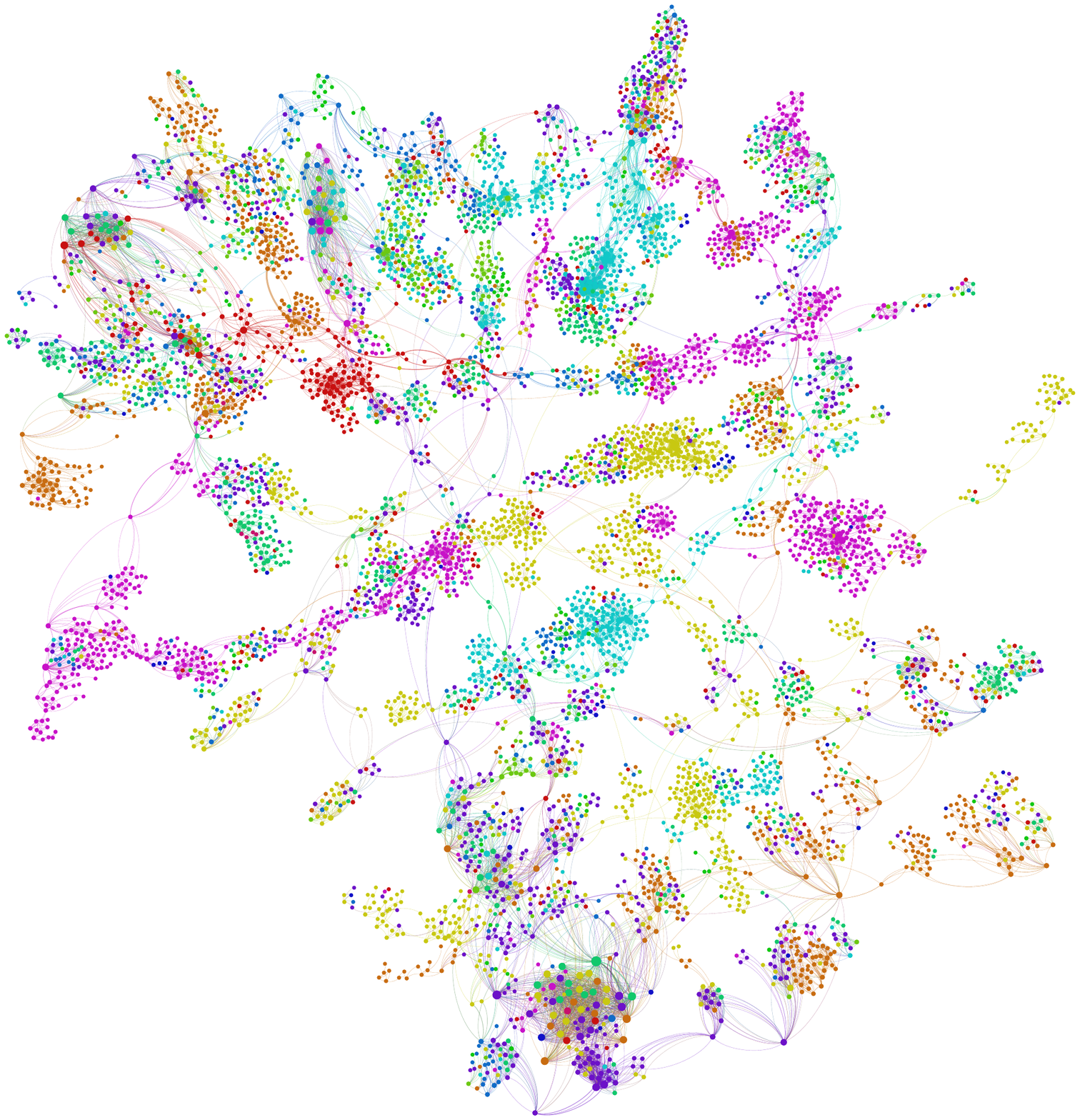}
\vspace{-2.6cm}
\caption{Name ethnicity on the giant component of the DBLP coauthorship network}
\label{fig-eth-coauthor-network}
\end{figure}

The visualized graph is shown in Figure~\ref{fig-eth-coauthor-network}, where color of a node indicates its name ethnicity and color of an edge is the average mix of its two nodes' color. It shows that most clusters appear in a single color, indicating name ethnicity as a strong homophily factor on academical social network.

\begin{figure*}
\centering
\includegraphics[width=0.95\textwidth]{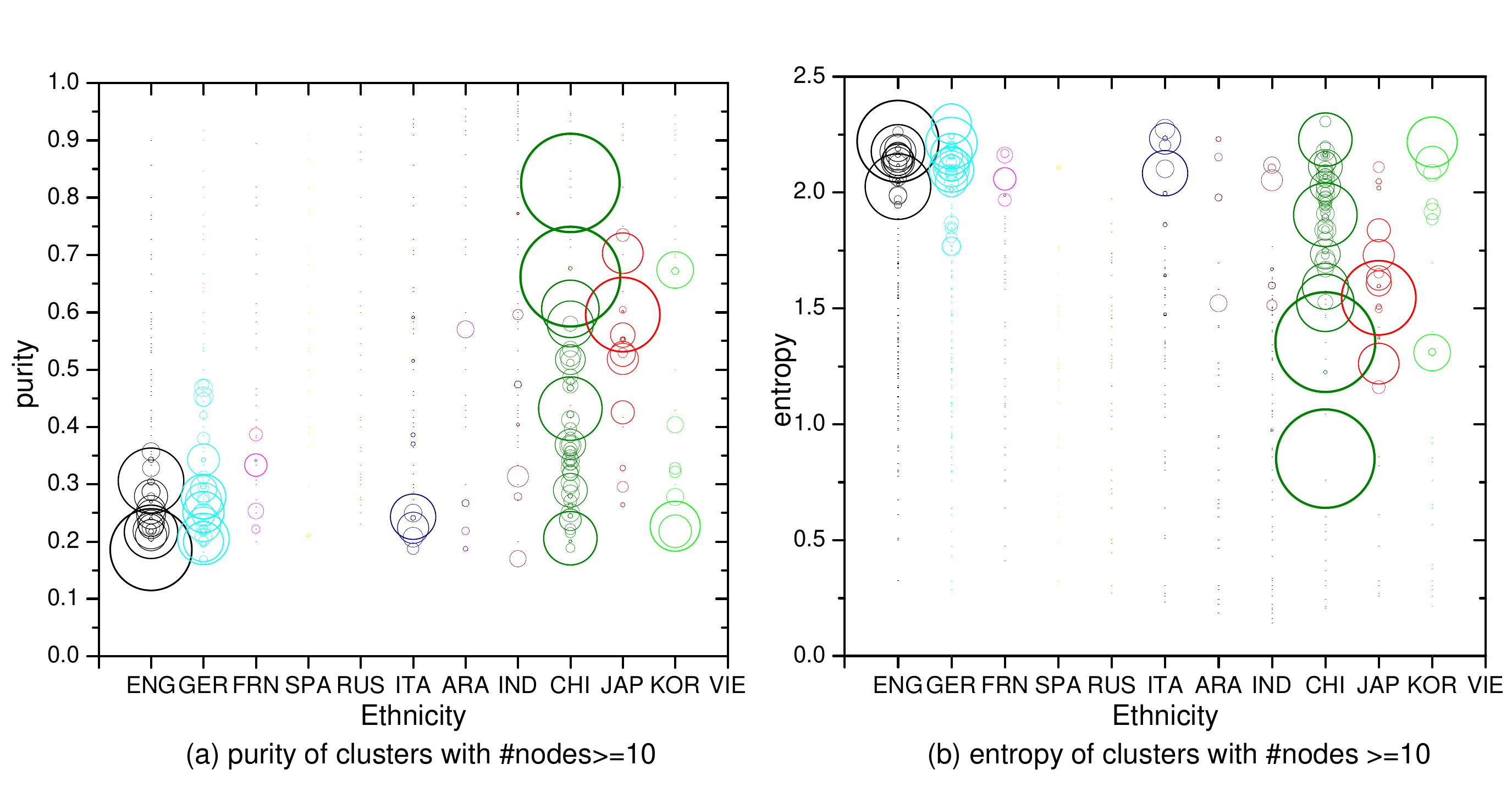}
\vspace{-3mm}
\caption{Purity and entropy of clusters with size >=10 on the DBLP coauthorship network}
\label{fig-purity}
\end{figure*}

To quantitatively study the effect of name ethnicity on coauthor network, two metrics, namely purity and entropy, are used together to measure the ethnic purity and diversity of a cluster. Suppose the name ethnicity set $E={e_i}$ and a cluster $c$ has a probability $p(e_i)$ to be labeled as name ethnicity $e_i$ ($\sum_i p(e_i) = 1$), which can be estimated by the maximum percentage of name ethnicity. The purity of a cluster is then defined as the maximum probability it can be assigned to a name ethnicity, i.e., $pry(c)=\max_i{p(e_i)}$. And the entropy~\cite{Claude:A}, to measure the diversity of a cluster, is given by $H(c) = -\sum_i p(e_i)\cdot log(p(e_i))$, where $p(e_i)$ is the probability that cluster $c$ should be labeled as name ethnicity $e_i$. For example, if in cluster $c$ there are 70\% ENG nodes and 30\% GER nodes, then $pry(c)$ is 0.7 (of being ENG) and $H(c)=-(0.7*log(0.7)+0.3*log(0.3))=0.61$.
The purity and entropy of all the clusters with number of nodes >= 10 is presented in Figure~\ref{fig-purity}. A cluster with purity of being name ethnicity $e_i$ is mapped to the x-axis of $e_i$ and the y-axis of the purity value. The size of a bubble indicates the size of a cluster.

\begin{figure}
\centering
\vspace{-4.2cm}
\includegraphics[width=0.5\textwidth]{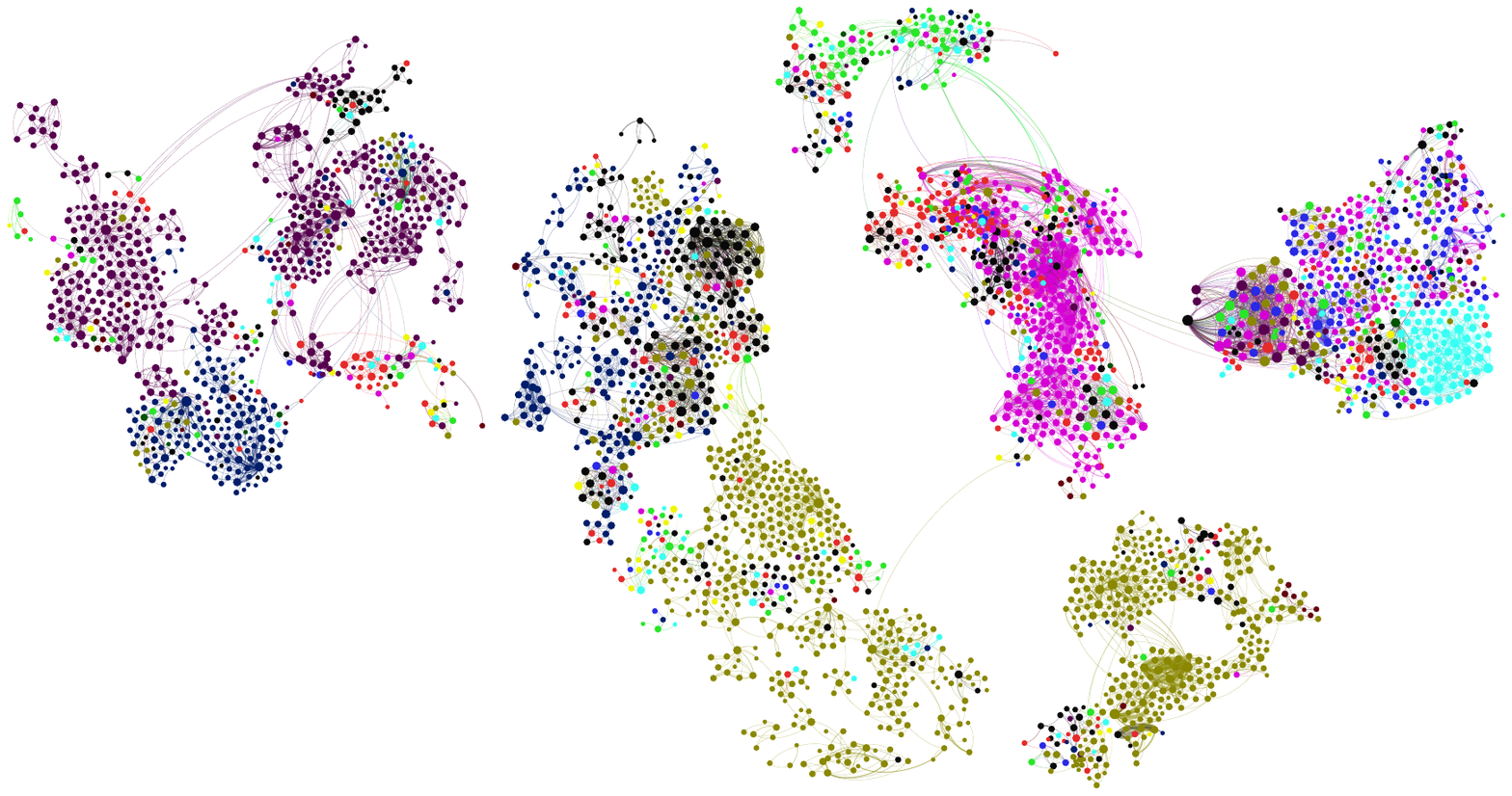}
\vspace{-4.8cm}
\caption{Name ethnicity on top 5 high purity clusters on the DBLP coauthorship network}
\label{fig-eth-community}
\end{figure}

Among all the 727 clusters, only 9 clusters have their purity smaller than 0.2 (>0.17) and only 4 clusters have their entropy slightly larger than the global entropy 2.24 ($p(e_i)$ is calculated by the percentage of $e_i$ on the whole network), which demonstrates the strong effect of name ethnicity on forming communities. As an illustrating example, the top 5 high purity clusters are shown in Figure~\ref{fig-eth-community}, where nodes/names of the same name ethnicity are in the same color (coordinated with colors used in Figure~\ref{fig-single-eth-contrib}). We can see that those clusters or their inside sub-clusters tend to be in the same color.
Moreover, Figure~\ref{fig-purity} clearly shows this effect differs in different ethnicities. For example, CHI and JAP tend to have larger high purity clusters than do others, indicating name ethnicity is a stronger homophily factor for CHI and JAP, or CHI and JAP are more likely to have intra-name ethnicity collaboration. ENG and GER tend to have larger high diversity clusters, suggesting their coauthorships are more open.

Our findings demonstrate the importance of name ethnicity as a homophily factor in research collaboration, which was not fully studied by existing works~\cite{SocialTie,Socialinfluenceanalysis,CollabSeer,Community:Deng,Groupformation}.
We believe name ethnicity could be used as a strong feature for social network analysis and modeling, especially for user modeling, social tie inference, and social recommendation.

\subsection{The Evolution of Collaboration}
The name ethnicity on coauthor network has been shown as a strong factor for shaping the formation of communities, especially for Asian ethnicities such JAP and CHI. However, it does not show the collaboration preference (as well as collaboration strength) among different ethnic groups and how do they evolve, thus give little hint to the evolution of ``European-Asian'' shift phenomenon.

We define $CS(e_i, e_j)$, the collaboration strength of two ethnic groups $e_i$ and $e_j$, as the number of their coauthored publications. An alternative to measure the collaboration strength is the total number of coauthorships between two ethnicities. The results are consistent with the publication centric measurement. For a paper with $K (>1)$ authors, $1/C(K,2)=2/K(K-1)$ paper is counted for every author pair. Hence, a paper with 2 ENG and 1 GER authors will be considered as $1/3$ ENG-ENG paper, $2/3$ ENG-GER paper while a paper with 3 ENG is 1 ENG-ENG paper. To give better observation on a name ethnicity's collaboration preference to other ethnicities, the normalized collaboration strength is used to measure the relative importance of $e_i$ to $e_j$ in collaboration, defined as the ratio of collaboration strength between $e_i$ and $e_j$ to the total collaboration strength of $e_j$.
\begin{equation}
NCS(e_i, e_j) = CS(e_i, e_j)/\sum_i{CS(e_i, e_j)}
\end{equation}
Note that $CS(e_i, e_j)$ is symmetric while $NCS(e_i, e_j)$ is symmetric, i.e., $CS(e_i, e_j)=CS(e_j, e_i)$ while $NCS(e_i, e_j)\neq NCS(e_j, e_i)$. For example, it is easy to infer that \\$NCS(\text{ENG}, \text{VIE}) > NCS(\text{VIE}, \text{ENG})$ since ENG contributes more to VIE's collaboration than does VIE contribute to ENG's.

\begin{figure*}
\centering
\includegraphics[width=\textwidth]{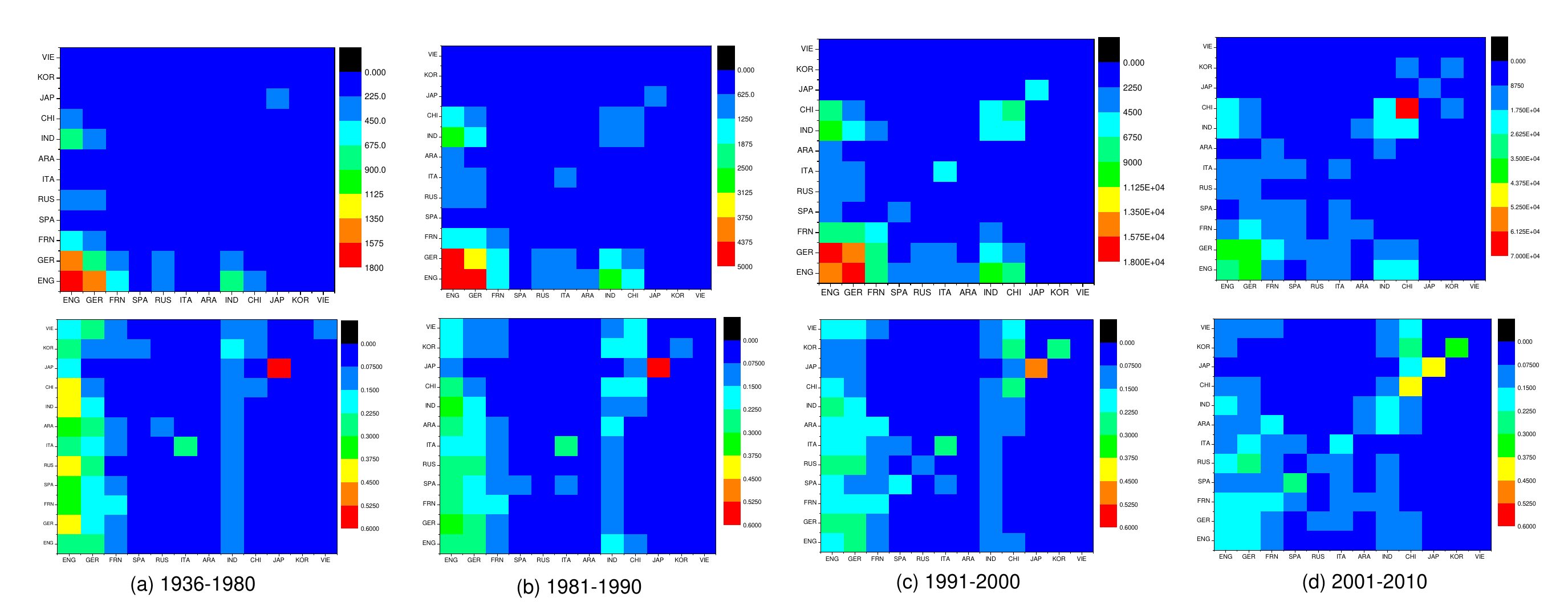}
\vspace{-6mm}
\caption{The evolution of collaboration strength among different name ethnicities from 1936 to 2010.}
\label{fig-eth-collab}
\end{figure*}

The evolution of collaboration strength of all name ethnicity pairs are shown by Figure~\ref{fig-eth-collab}, in 4 different periods, (a) 1936-1980, (b) 1981-1990, (c) 1991-2000 and (d) 2000-2010. In each period, sub-figure in the first row depicts the absolute collaboration strength while sub-figure in the second row presents the corresponding normalized collaboration strength. Since the value of $CS$ varies too much in different periods while the $NCS\in[0,1]$, we use different scales for coloring the four sub-figures in the first row while the same scale for coloring the four sub-figures in the second row.

First, clear collaboration preferences exist in collaboration among different ethnicities. Indian collaborates more with European name ethnicities while other Asian ethnicities such as Japanese, Korean and Chinese tend to have more intra-name ethnicity collaboration or collaborate more with other Asian ethnicities, which is consistent with the previous finding shown by Figure~\ref{fig-purity}. It is interesting to note that JAP is the name ethnicity with the strongest intra-name ethnicity all the time.

Second, for each ethnic group, both intra-name ethnicity and inter-name ethnicity collaborations are steadily increasing while the intra-name ethnicity dominates and increases much faster. This is clearly demonstrated by the evidence that the diagonal is getting stronger over time.

Third, ethnicities in different developing phases have their own developing trajectories: better developed ethnicities like English or German stably increase their intra-name ethnicity and inter-name ethnicity collaborations while developing ethnicities first collaborate with those better developed ones and then strengthen their intra-name ethnicity collaborations. This directly leads to the phenomenon of ``European-Asian'' shift discussed below.

Fourth, the normalized collaboration strength of ENG and GER are gradually decreasing while that of IND and CHI are gradually increasing. And it is worth to note that CHI has replace ENG's position as the most important collaborator to Asian ethnicities such as KOR, JAP and VIE.

We believe these findings could provide suggestive insight for policy making towards research and education and hope they can also give a better vision of the evolution of the computer science field and shed some light on other social and scientific behavior trends.
\section{Related Work}
To the best of our knowledge, there is no previous work on the evolution of science based on name ethnicity classification. We review the work on name ethnicity classification and its related applications.

Name-based ethnicity classification has gathered much interest especially in biomedical research~\cite{Coldman:classification,FiscellaUse,Mateos,Gill:Limitations}. However, most previous works in name ethnicity classification focus on binary classification (whether a name belongs to a name ethnicity or not) or to a small number of specific name ethnicity groups, while little effort has been made to develop the multi-faceted classifier capable of discriminating names among a large number of ethnicities except~\cite{Ambekar:opensources,ePluribus}.
The most common used method in name ethnicity classification is to compare to existing name lists, relying on name frequency. For example, Goldman et al. use a simple probabilistic method based on full name lists to identify people with Chinese name ethnicity~\cite{Coldman:classification}. Gill et al. use surname analysis together with location information to better infer name ethnicity from names~\cite{Gill:Limitations}. More recently, Chang et al. train a graphical generative model based on US Census names to infer ethnicities of Facebook users from names and studied the interactions between ethnic groups~\cite{ePluribus}. They found that different ethnic groups relate to one another in an assortative manner and that these groups differ in profiles.

More recently, Ambekar et al. proposed a method for cultivating open data sources such as Wikipedia for generating name-ethnicity dictionary~\cite{Ambekar:opensources} and used a Hidden Markov Model to model probability transitions between character sequences in first names, middle names and last names respectively. To the best of our knowledge, their work is the first to take advantage of open data sources and syntactic information in names for name ethnicity classification. This work is the most similar to our name ethnicity classifier.

Besides using name ethnicity, there are some works using real ethnicity data from third parties. For example, Ginther et al. investigated the association between applicants' ethnicity and the probability of receiving an award by using data from the NIH IMPAC II grant database and the Thomson Reuters Web of Science~\cite{eth:11}. They found the black applicants are underestimated to be awarded NIH research funding compared with the white. However, those real ethnicity data is usually not publicly available and hard to obtain.

In addition, Menezes et al. gave a geographical analysis of knowledge production in Computer Science based on DBLP coauthorship networks in different regions of the world~\cite{geographical:Menezes}, showing different collaboration patterns. In scientometrics~\cite{Luukkonen:Understanding,W:National}, international scientific collaboration is studied using bibliometric methods via special national databases which can provide accurate institute and nation information of authors, but usually based on small set of publications over a short range of time. Comparing to nationality, ethnicity might provide a new feature and give different insights for understanding scientific behaviors.

\section{Conclusion and Future Work}
We presented a systematic study on the impact of name ethnicity in computer science research world. The ethnicity identification is completely based on name information derived from our name ethnicity classifier trained on a large set of Wikipedia names. The population dynamics and the evolution of ethnicity composition and scientific output showed the history and evolution of different name ethnicities, not only in all of computers science, but also in various research communities.

At a higher level, the goal of this work is to understand different name ethnicities in the computer science discipline and provide a foundation for reasoning about the evolution of name ethnicities in different research communities and the whole of science and academia. Our work reveals a set of interesting findings and thus also opens up a range of questions to not only the computer science community, but also social science, political science and economics~\cite{Teasley:Scientific,Norman:Scientific,Kerr:Ethnic}. In particular, it would be interesting to investigate how the ethnicity as a homophily factor could remodel social link prediction and collaboration recommendation, as well as how the status of different ethnicities could provide insights or suggestions to policy making on education and immigration issues. Furthermore, we are interested to see if large scale studies in other areas such as MEDLINE/PubMed would yield similar or conflicting trends.

\bibliographystyle{abbrv}
\bibliography{ethnicityCS2arxiv}
\end{document}